\documentclass[12pt]{article}
\usepackage{hyperref}
\usepackage{cite}
\usepackage{color}
\usepackage{footnote}
\makesavenoteenv{minipage}
\renewcommand{\title}[1]{%
    \bigskip%
    \begin{center}%
    \Large\bf #1%
    \end{center}%
    \vskip .2in}

\renewcommand{\author}[1]{%
    {\begin{center}
    #1
    \end{center}}}
\newcommand{\address}[1]{\vspace{-1.7em}\vspace{0pt}
    {\begin{center}
    \it #1 
    \end{center}}}

\begin{document}

\title{Generalized supersymmetry and sigma models}

\author
{\bf 
Rabin Banerjee\footnote{E-mail: rabin@bose.res.in},
Sudhaker Upadhyay
\footnote{E-mail:  sudhakerupadhyay@gmail.com;\ sudhaker@bose.res.in }}


\address{ S. N. Bose National Centre 
for Basic Sciences, JD Block, \\ Sector III, Salt Lake City, Kolkata -700 098, India }

\begin{abstract}
 In this paper, we discuss the generalizations of exact supersymmetries present in the
 supersymmetrized sigma models.
  These generalizations are made by making the supersymmetric
 transformation parameter  field-dependent.
 Remarkably, the supersymmetric  effective actions  
 emerge naturally through the Jacobian associated with
 the generalized supersymmetry transformations.
 We   explicitly demonstrate these  for two different
 supersymmetric sigma models, namely, one dimensional sigma model and topological 
 sigma model for hyperinstantons on quaternionic manifold. 
 
\end{abstract}

\section{Introduction}

Supersymmetry is one of the most important concepts in modern theoretical physics, especially, in the search of unified theories beyond the standard model \cite{hub}. In particle physics, for example,  the supersymmetric standard model predicts the existence of a
superpartner for every particle in the standard model. However, theoretical 
understanding of supersymmetry is quite far from complete.
To examine the non-perturbative aspects of supersymmetric standard model, the utilization
of the so-called space-time lattice simulation method is quite obscure as the theory involves 
many different scales. Supersymmetry is also relevant  in string theories
also though it is quite far from the real experimental world. The advantage of superstring theories (those string models which also incorporate supersymmetry) is that
 it does not predict the existence of a bad behaving particle called the Tachyon. 
 In particle theory, supersymmetry  finds a way to stabilize the hierarchy between the unification scale and the electroweak scale  or the Higgs boson mass.
 Supersymmetry models are also considered as a natural dark matter candidate \cite{jung}.

Since it encompasses both  theoretical and phenomenological interests, some 
serious attempts have been made to study supersymmetric theories \cite{kiku, kat}.  But these attempts encountered some problems like supersymmetry breaking as well as fine-tuning.
Recent developments have been made in the construction of lattice actions which
possess a subset of the supersymmetries of the continuum theory and have
a Poincar\'{e} invariant continuum limit \cite{car}. The presence of the exact supersymmetry  provides a way to obtain the continuum limit with no
fine tuning or fine tuning much less than conventional lattice constructions.
 The remarkable feature  of presence of exact
supersymmetry is that it reduces and in some cases eliminates the need for fine tuning
to achieve a continuum limit invariant under the full supersymmetry of the
target theory \cite{car,cat,cat1}.
However, the construction of the  supersymmetric non-linear sigma model with $O(N)$ 
target manifold was first made by Witten \cite{wit} and then by
P. Di Vecchia and S. Ferrara \cite{vec} which describe the spontaneous breaking of
chiral symmetry and the dynamical generation of particle masses \cite{a,b,c,d}.
 Subsequently, the geometric
interpretation of  supersymmetric sigma models were classified 
in terms of BRST operator \cite{alva, bau}.
These sigma models are described by maps between a two-dimensional space called the world-sheet and some target space, taken
to be a manifold in this setting. 
The connections of supersymmetry and geometry  became more stronger after Witten's seminal construction of  the so--called topological twist \cite{witt}.
The motivation behind the twist is that in
a topological field theory one can compute certain physical quantities more easily than in the original theory, where we sometimes lack the tools to compute them exactly. 
The topological sigma models in four dimensions are also used in the study of triholomorphic maps
on hyperK\"{a}hler manifolds \cite{fre}. 
A naive discussion of gauge invariant topological field theory is presented in BRST-BV 
framework \cite{jmf}.

 On the other hand, generalization of BRST transformation
 by making the infinitesimal parameter finite and field-dependent was first developed 
 in  \cite{bm} which is known as
 finite field-dependent BRST (FFBRST) transformation.
Such generalizations have found various applications in gauge field theories 
as well as in M-theory 
 \cite{bm,mru,sdj1,rb,susk,jog,sb1,smm,sudd,sud,fs,rbs}. 
 However this generalization of BRST technique has, as yet, not been done 
 for supersymmetry. Considering the deep connection between BRST
 and supersymmetry we feel that this is a glaring omission. 
 
 The aim of the present
 paper is to
investigate the features of generalized supersymmetry in the framework of FFBRST formulation. 
Specifically, we consider   supersymmetric sigma model
and supersymmetric topological sigma model  in a gauge invariant framework. 
Further, we discuss the generalizations of supersymmetries present in the theory in a detailed way. 
These generalizations  are made by making the infinitesimal transformation parameter   finite and field-dependent.
Further, we stress the significant features of this
generalized supersymmetry. For instance, we find that while
the effective actions are invariant under generalized supersymmetry,  the
measures of path integrals are not. The obvious reason for this is that the
path integral measure changes non-trivially.
This  non-trivial Jacobian plays a significant role
in the formation of supersymmetric actions for sigma models. 
 We show that the path integral measure under generalized supersymmetry transformation with some specific  choices of parameter reproduces   exactly 
the same effective actions as the original theories.  In other words, 
the supersymmetric actions proposed in the literature \cite{cat, fre} may be systematically obtained
within the framework of FFBRST transformations.
 We analyse results in one dimensional supersymmetric  sigma model
 and in supersymmetric topological sigma model where the gauge-fixing is
provided by the triholomorphic instanton condition.  Even though we establish
the results with the help of specific examples but this works for a general
supersymmetric invariant theory.
 
The paper is organized in four sections.
First, we provide the mechanism to    generalize the supersymmetry
in FFBRST framework in section 2. In section 3, which is the main section 
of the paper, we show that the Jacobians of the functional measures  for 
FFBRST transformations with judicious choices of the transformation
parameters naturally yield  the supersymmetric actions 
for sigma models.
 We draw concluding remarks in the last section.
 \section{Generalized supersymmetric BRST transformation}
In this section,  we briefly review the generalized supersymmetric BRST formulation of pure gauge theories by making the infinitesimal
parameter finite and field-dependent. It is a supersymmetric generalization
of finite field dependent BRST (FFBRST)  transformation originally
advocated in \cite{bm} for the non-supersymmetric cases.
 We first present the general  methodology for the standard  Maxwell theory in Euclidean space-time. 
For this purpose, let us start by defining the
partition function for BRST invariant  Maxwell theory in four dimensions as following
\begin{eqnarray}
Z_{M} =\int {\cal D} A_\mu {\cal D}c {\cal D}\bar c{\cal D} B e^{-S_M},
\end{eqnarray}
where the effective action $S_M$ in Lorentz gauge is defined by
\begin{eqnarray}
S_M =\int d^4 x\left[ -\frac{1}{4} F_{\mu\nu}F^{\mu\nu} +\frac{1}{2 } B^2 -B\partial_\mu A
^\mu  +
\partial_\mu \bar c \partial^\mu c\right].\label{sm}
\end{eqnarray}
Here $B$, $c$ and $\bar c$ are Nakanishi-Lautrup, ghost  and anti-ghost fields
respectively.
This effective action as well as the partition function are invariant under usual
BRST transformations
\begin{eqnarray}
\delta_b A_\mu (x)&=& \partial_\mu c(x)\ \delta\Lambda,\nonumber\\
\delta_b c(x) &=& 0,\nonumber\\
\delta_b \bar c (x)&=& B(x)\ \delta\Lambda,\nonumber\\
\delta_b B(x) &=&0, 
\end{eqnarray}
where $ \delta\Lambda$
is an infinitesimal, anticommuting and global parameter.
Most of the features of the BRST transformation do not depend on whether 
the parameter $\delta\Lambda$  is (i) finite or infinitesimal, (ii) field-dependent or not, as long 
as it is anticommuting and space-time independent. These observations give us a freedom to 
generalize the BRST transformation by making the parameter, $\delta\Lambda$, finite and field-dependent without
 affecting its properties. To generalize such transformation we 
start 
by making the  infinitesimal parameter field-dependent with introduction of an arbitrary parameter $\kappa\ 
(0\leq \kappa\leq 1)$.
We allow the generic fields, $\Phi(x,\kappa)$, to depend on  $\kappa$  in such a way that $\Phi(x,\kappa =0)=\Phi(x)$ and $\Phi(x,\kappa 
=1)=\Phi^\prime(x)$, the transformed field.

The usual infinitesimal  transformation, thus can be written generically as \cite{bm}
\begin{eqnarray}
\frac{ dA_\mu(x,\kappa)}{d\kappa} &=& \partial_\mu c(x)\ \Theta^\prime [\Phi (x,\kappa ) ],\nonumber\\
\frac{ dc(x,\kappa)}{d\kappa} &=& 0,\nonumber\\
\frac{ d\bar c(x,\kappa)}{d\kappa} &=& B(x)\ \Theta^\prime [\Phi (x,\kappa ) ],\nonumber\\
\frac{d B(x,\kappa)}{d\kappa} &=&0, \label{dif}
\end{eqnarray}
where the $\Theta^\prime [\Phi (x,\kappa ) ]$ is the infinitesimal but field-dependent parameter.
The FFBRST transformation ($\delta_f$) then can be 
constructed by integrating such infinitesimal transformation from $\kappa =0$ to $\kappa= 1$,  as
\begin{eqnarray}
\delta_f A_\mu (x)&=& A_\mu  (x,\kappa =1)-A_\mu (x,\kappa=0)=\partial_\mu c(x)\ \Theta[\Phi(x) ],\nonumber\\
\delta_f c (x)&=& c (x,\kappa =1)-c(x,\kappa=0)= 0,\nonumber\\
\delta_f \bar c (x)&=& \bar c(x,\kappa =1)-\bar c(x,\kappa=0)=B(x)\ \Theta[\Phi(x) ],\nonumber\\
\delta_f B (x)&=& B(x,\kappa =1)-B(x,\kappa=0)=0, 
\end{eqnarray}
where \cite{bm}
\begin{equation}
\Theta[\Phi(x)]=\int_0^1 d\kappa^\prime\Theta^\prime [\Phi(x,\kappa^\prime)],
\label{fin}
\end{equation}
 is the finite field-dependent parameter. 
Such a generalized transformation with finite field-dependent
 parameter is a symmetry  of the effective action $S_M$,
 i.e.,
  \begin{eqnarray}
 \delta_f S_M =(s_b S_M)\Theta =0,
 \end{eqnarray}
   where $s_b$ is Slavnov variation. 
 Let us explicitly show the invariance of
the Maxwell term. Under the transformations (5), the Maxwell pieces
changes as,

\begin{eqnarray}
\label{new}
\delta_f (F_{\mu\nu}F^{\mu\nu})&=& 4F_{\mu\nu}\delta_f \partial^\mu A^\nu,
\\
\nonumber
&=&4F_{\mu\nu} \partial^\mu \left[\partial^\nu c \Theta\right],\\
\nonumber
&=&0.
\end{eqnarray}
Since the FFBRST parameter $\Theta$ is spacetime independent the
derivative acts only on the variable $c$. By symmetry this term vanishes.
Hence the Maxwell piece remains invariant.  
  Although the action remains invariant, the functional measure is not invariant under such a transformation as the 
Grassmann parameter is field-dependent in nature. 
The Jacobian, $J(\kappa )$, of path integral measure changes nontrivially and can be
replaced  as \cite{bm} 
\begin{equation}
J(\kappa )\longmapsto e^{-S_1 [\Phi(x,\kappa) ]},
\end{equation}
 if and only if the following condition is satisfied as we do not
 want any numerical change in the path integral measure \cite{bm}
\begin{eqnarray}
 \int {\cal{D}}\Phi (x) \;  \left [  \frac{d}{d\kappa}\ln J(\kappa)+\frac
{dS_1 [\Phi (x,\kappa )]}{d\kappa}\right ] e^{ - S_1 [\Phi(x,\kappa) ]  }=0, \label{mcond}
\end{eqnarray}
where $ S_1[\Phi ]$ is some local functional of fields satisfying an
initial boundary condition
\begin{equation}
S_1[\Phi]_{\kappa=0} =0.\label{ini}
\end{equation}

Furthermore, the infinitesimal change of the logarithm of $J(\kappa)$ can be calculated from the formula \cite{bm}:
\begin{equation}
 \frac{d}{d\kappa}\ln J(\kappa)=-\int d^4x\left [ \partial_\mu c(x)\frac{
\partial\Theta^\prime [\Phi (x,\kappa )]}{\partial A_\mu (x,\kappa )}- B(x)\frac{
\partial\Theta^\prime [\Phi (x,\kappa )]}{\partial\bar c (x,\kappa )}\right]. \label{i}
\end{equation}
 For a particular choice of   $\Theta^\prime [\Phi (x,\kappa )]$ given by,
 \begin{equation}
\Theta^\prime [\Phi (x,\kappa )]=-\int  d^4 x\ \bar c [\partial_\mu A^\mu (x,\kappa )-\eta_\mu A^\mu (x,\kappa)],
\end{equation}
the expression in (\ref{i}) reduces to
\begin{eqnarray}
 \frac{d}{d\kappa}\ln J(\kappa)&=& \int d^4x\left [- \partial_\mu c  \partial^\mu \bar c
 - \partial_\mu c  \eta^\mu \bar c- B\partial_\mu A^\mu +B\eta_\mu A^\mu
  \right],\nonumber\\
 &=& \int d^4x\left [\partial_\mu \bar c  \partial^\mu  c
 + \eta^\mu \bar c\partial_\mu c  - B\partial_\mu A^\mu +B\eta_\mu A^\mu
  \right]. \label{j}
\end{eqnarray}
Now, an ansatz for the functional $S_1[\Phi ]$ is taken as
\begin{eqnarray}
S_1 =\int d^4x\left [  \zeta_1(\kappa) B\partial_\mu A^\mu +\zeta_2(\kappa) B\eta_\mu A^\mu +
 \zeta_3(\kappa) \partial_\mu \bar c  \partial^\mu  c
 +\zeta_4(\kappa) \eta^\mu \bar c\partial_\mu c  \right],
\end{eqnarray}
where $\zeta_i (i=1,2,3,4)$ are arbitrary constant parameters
constrained by 
\begin{equation}
\zeta_i(\kappa=0)=0,\label{bo}
\end{equation}
so that the requirement (\ref{ini}) holds.

To satisfy the essential condition (\ref{mcond}),  we calculate the 
$dS_1/d\kappa$ by 
employing   (\ref{dif}) as follows:
\begin{eqnarray}
\frac{dS_1}{d\kappa}& =&\int d^4x\left [  \frac{d\zeta_1}{d\kappa}  B\partial_\mu A^\mu +\frac{d\zeta_2}{d\kappa} B\eta_\mu A^\mu +\frac{d\zeta_3}{d\kappa}   \partial_\mu \bar c  \partial^\mu  c
 +\frac{d\zeta_4}{d\kappa}  \eta^\mu \bar c\partial_\mu c \right.\nonumber\\
 &+&\left. (\zeta_1+\zeta_3)  B  (\partial_\mu\partial^\mu  c)\Theta'
 + (\zeta_2  -\zeta_4)  B  (\eta_\mu\partial^\mu  c)\Theta'
  \right].\label{s}
\end{eqnarray}
The condition (\ref{mcond}) along with Eqs. (\ref{j}) and (\ref{s}) leads to
\begin{eqnarray}
&  &\int d^4x\left [ \left( \frac{d\zeta_1}{d\kappa}-1\right)  B\partial_\mu A^\mu + \left( \frac{d\zeta_2}{d\kappa}+1\right) B\eta_\mu A^\mu + \left( \frac{d\zeta_3}{d\kappa}+1\right) \partial_\mu \bar c  \partial^\mu  c
 \right.\nonumber\\
 &&+ \left( \frac{d\zeta_4}{d\kappa}+1\right) \eta^\mu \bar c\partial_\mu c +\left. (\zeta_1+\zeta_3)  B  (\partial_\mu\partial^\mu  c)\Theta'
 + (\zeta_2  -\zeta_4)  B  (\eta_\mu\partial^\mu  c)\Theta'
  \right] =0.
\end{eqnarray}
The last two non-local ($\Theta'$-dependent) terms disappear from
the above equation for
$(\zeta_1+\zeta_3) =(\zeta_2  -\zeta_4)=0$. However, the disappearance of local terms
yields the following
differential equations
\begin{eqnarray}
 \frac{d\zeta_1}{d\kappa}-1&=&0,\ \ \ \frac{d\zeta_2}{d\kappa}+1=0, \nonumber\\
  \frac{d\zeta_3}{d\kappa}+1&=&0,\ \ \  \frac{d\zeta_4}{d\kappa}+1=0. 
\end{eqnarray}
The solutions of the  above equations satisfying the boundary conditions
(\ref{bo})
are
\begin{eqnarray}
\zeta_1 =\kappa,\ \ \zeta_2 =-\kappa,\ \ \zeta_3 =-\kappa,\ \ \zeta_4=-\kappa.
\end{eqnarray}
With these identifications, the
functional $S_1[\Phi(x,\kappa),\kappa]$ has the form
\begin{eqnarray}
S_1[\Phi(x,\kappa),\kappa] =\int d^4x\left [  \kappa B\partial_\mu A^\mu - \kappa  B\eta_\mu A^\mu -\kappa  \partial_\mu \bar c  \partial^\mu  c
 -\kappa  \eta^\mu \bar c\partial_\mu c  \right],
\end{eqnarray}
which vanishes at $\kappa =0$.
Now, by adding this $S_1[\Phi(x,\kappa),\kappa]$ to $S_M$ given in (\ref{sm}), we obtain
\begin{eqnarray}
S_M +S_1[\Phi(x,\kappa),\kappa]&= &\int d^4 x\left[ -\frac{1}{4} F_{\mu\nu}F^{\mu\nu} +\frac{1}{2 } B^2 -(1-\kappa) B\partial_\mu A
^\mu  +(1-\kappa)
\partial_\mu \bar c \partial^\mu c\right.\nonumber\\
&-&\left.   \kappa  B\eta_\mu A^\mu -\kappa  \eta^\mu \bar c\partial_\mu c \right].\label{mix}
\end{eqnarray}
At $\kappa =0$, the above expression reduces to
\begin{eqnarray}
S_M +S_1[\Phi(x,0),0]&= &\int d^4 x\left[ -\frac{1}{4} F_{\mu\nu}F^{\mu\nu} +\frac{1}{2 } B^2 -  B\partial_\mu A
^\mu  + 
\partial_\mu \bar c \partial^\mu c  \right],
\end{eqnarray}
which is the original theory in Lorentz gauge.
However, at $\kappa =1$ (under FFBRST transformation) the  expression (\ref{mix}) within a 
functional integration effectively
reduces to the Maxwell action in axial gauge as given below
\begin{eqnarray}
S_M +S_1[\Phi(x,1),1] &= &\int d^4 x\left[ -\frac{1}{4} F_{\mu\nu}F^{\mu\nu} +\frac{1}{2 } B^2    - B\eta_\mu A^\mu -  \eta^\mu \bar c\partial_\mu c \right].
\end{eqnarray}
This shows that the FFBRST formulation is able to connect two different
gauge fixed versions of the maxwell theory. Incidentally, this
was the original motivation for developing the FFBRST transformation.

A natural question that arises in this context is the possibility 
of generating the action itself throgh FFBRST formulation. To answer this question it is useful to ponder on the structure of the 
jacobian (13). This involves terms that are subsequently interpreted as a combination of gauge fixing and ghost terms. Such
combinations, which are BRST exact, appropriately modify the structure to connect the Maxwell theory in distinct gauges. It is
clear, therefore, that since the jacobian is BRST exact, this by itself would fail to generate the Maxwell action simply
because it is not BRST exact. Hence, in order for the jacobian to reproduce the whole action, that particular action must be BRST exact.
 Such a possibility occurs for the supersymmetric sigma models. To implement
 these notions, therefore, it is essential to first extend the FFBRST formulation to include  supersymmetry.

To generalize the FFBRST formulation for supersymmetric transformation, let us write the 
usual supersymmetric transformation for a collective field $\Phi$  of 
sigma models,
\begin{equation}
\delta  \Phi = {\cal R}[\Phi] \xi,
\end{equation}
  where ${\cal R}[\Phi]$ is supersymmetric variation of $\Phi$
  and $\xi$ is infinitesimal parameter of transformation. This observation  gives us a freedom to 
generalize the supersymmetry transformation in the same fashion as discussed above by making the parameter, $\xi$, finite and field-dependent.  We first define the  infinitesimal field-dependent transformation as
\begin{equation}
\frac{d\Phi(\sigma,\kappa)}{d\kappa}={\cal R} [\Phi (\sigma,\kappa ) ]\Theta^\prime [\Phi (\sigma,\kappa ) ],
\label{diff}
\end{equation}
where the $\Theta^\prime [\Phi (\sigma,\kappa ) ]$ is an infinitesimal  field-dependent parameter and $\sigma$ is a parameter which parametrizes the base space
of sigma models.
The generalized supersymmetry ($\delta_g$) with the finite field-dependent parameter then can be 
obtained by integrating the above transformation from $\kappa =0$ to $\kappa= 1$, as follows:
 \begin{equation}
\delta_g \Phi(\sigma)\equiv \Phi (\sigma,\kappa =1)-\Phi(\sigma,\kappa=0)={\cal R}[\Phi(\sigma) ]\Theta[\Phi(\sigma) ],
\end{equation}
where 
$\Theta[\Phi(\sigma)]$
 is the finite field-dependent parameter constructed from its infinitesimal version using
 (\ref{fin}) written in base space. 
Under such generalized supersymmetry transformation with finite field-dependent
 parameter  the measure of partition function will  not be invariant
 and will contribute some non-trivial terms to the partition function in general.

The Jacobian of the path integral measure $({\cal D}\Phi)$ in the functional 
integral for such transformations is then evaluated for some 
particular choices of the finite field-dependent parameter, $\Theta[\Phi(\sigma)]$, as
\begin{eqnarray}
{\cal D}\Phi^\prime &=&J(
\kappa) {\cal D}\Phi(\kappa).\label{jacob}
\end{eqnarray}
Now we  replace the Jacobian $J(\kappa )$ of the path integral measure  as   
\begin{equation}
J(\kappa )\longmapsto e^{-S [\Phi(\sigma,\kappa) ]},\label{js}
\end{equation}
 by paying the  cost  that the given condition (\ref{mcond}) must be satisfied 
where $ S[\Phi ]$ is some local functional of fields satisfying 
initial boundary condition given in (\ref{ini}).
 
Moreover, the infinitesimal change in Jacobian, $J(\kappa)$, as before,
\begin{equation}
 \frac{d}{d\kappa}\ln J(\kappa)=-\int d^m\sigma\left [\pm\sum_i {\cal R}[\Phi^i(\sigma )]\frac{
\partial\Theta^\prime [\Phi (\sigma,\kappa )]}{\partial\Phi^i (\sigma,\kappa )}\right],\label{jac}
\end{equation}
where, for bosonic fields,  $+$ sign is used and for fermionic fields,
$-$ sign is used.

\section{ Sigma models}
In this section, we will use the supersymmetric FFBRST mechanism 
to generate the actions for two distinct sigma models. First, we discuss
 the  sigma model on a curved target space and then a   
topological sigma model on quaternionic manifolds.  
\subsection{Sigma model on a curved target space} 
To discuss the sigma model, let us start 
 by considering the real bosonic field  $\phi^i (\sigma)$
corresponding to coordinates  on a Riemannian target manifold with metric $g_{ij}$
where the coordinate $\sigma$ parametrizes the one dimensional 
base space. This theory is supersymmetrized by considering two more
 real fermionic fields $\psi_i (\sigma)$ and $\eta_i (\sigma)$
and one  Lagrange  multiplier (bosonic) field $B_i (\sigma)$.
Now,  the  infinitesimal supersymmetry transformations  
parametrized by a global Grassmann parameter $\xi$ are given by \cite{cat}
\begin{eqnarray}
\delta  \phi^i &=& -\psi^i \xi,\nonumber\\
\delta \psi^i &=&0,\nonumber\\
\delta  \eta_i &=& \left(B_i -\eta_j\Gamma^j_{\ ik}\psi^k \right)\xi,\nonumber\\
\delta  B_i &=& -\left(B_j\Gamma^j_{\ ik}\psi^{k} -\frac{1}{2}\eta_j R^j_{\ ilk}\psi^l\psi^k \right)\xi,\nonumber\\
\delta  \Gamma^j_{\ ik} &=&\partial_m \Gamma^j_{\ ik}\psi^m \xi,\nonumber\\
\delta  R^j_{\ ilk} &=&\partial_m R^j_{\ ilk}\psi^m \xi,
\end{eqnarray}
where, in terms of
affine connection $\Gamma^j_{\ ik}$, the Riemannian curvature tensor $R^i_{\ jkl}$ is defined by:
\begin{eqnarray}
R^i_{\ jkl}= \partial_k \Gamma^i_{\ jl} -\partial_l \Gamma^i_{\ jk} +
\Gamma^i_{\ mk} \Gamma^m_{\ jl} -\Gamma^i_{\ ml} \Gamma^m_{\ jk}.
\end{eqnarray}
For any general fields $f(\sigma)$ and $g(\sigma)$, the supersymmetric operator  $\delta$ 
acts on the composite field   $f\cdot g$  as follows
 $(\delta f)\cdot g + f\cdot (\delta g)$.
With this definition, the nilpotency of operator $\delta$ ( i.e., $\delta^2=0$)   can be 
proved easily in the following manner:
\begin{eqnarray}
\delta^2 \phi^i &=&\delta \psi^i =0,\nonumber\\
\delta^2\eta_i &=& \delta  B_i - \delta\eta_j\Gamma^j_{\ ik}\psi^k 
 -\eta_j\delta\Gamma^j_{\ ik}\psi^k =0,\nonumber\\
 \delta^2  B_i &=& -\delta B_j\Gamma^j_{\ ik}\psi^{k}- B_j \delta\Gamma^j_{\ ik}\psi^{k} +\frac{1}{2}\delta\eta_j R^j_{\ ilk}\psi^l\psi^k  +\frac{1}{2}\eta_j \delta R^j_{\ ilk}\psi^l\psi^k =0,\nonumber\\
 \delta^2  \Gamma^j_{\ ik} &=&\partial_m \partial_n\Gamma^j_{\ ik}\psi^n\psi^m =0,\nonumber\\
\delta  R^j_{\ ilk} &=&\partial_m\partial_n R^j_{\ ilk}\psi^n\psi^m =0.
\end{eqnarray}
Now,  the supersymmetric action  for the sigma model in one dimension, which remains
invariant under the above fermion transformations, is given by \cite{cat}
\begin{eqnarray}
S &=&\alpha\int d\sigma \left[ B_iN^i(\phi) -\frac{1}{2}  
 g^{ij}B_iB_j-  \eta_i \nabla_k N^i \psi^k + \frac{1}{4}  R_{jlmk}\eta^j\eta^l\psi^m\psi^k\right],
\end{eqnarray}
where $N^i (\phi)$ denotes an arbitrary gauge-fixing condition for the bosonic
field $\phi^i$ and $\alpha$ is a coupling constant. Here we note that the
supersymmetric invariant observables do not depend on the choice of $\alpha$.
The symbol $\nabla_k$ indicates the general target space covariant derivative. For the sigma model the most convenient   gauge-fixing condition  is
\cite{cat},
\begin{equation}
N^i (\phi) =\frac{d\phi^i}{d\sigma}.\label{con}
\end{equation}
For this particular choice the
above action reduces to the form:
\begin{eqnarray}
S &=&\alpha\int d\sigma \left[  B_i\frac{d\phi^i}{d\sigma} -\frac{1}{2}  
 g^{ij}B_iB_j-  \eta_i \left( \frac{d\psi^i}{d\sigma}+\Gamma^i_{\ kj}\frac{d\phi^k}{d\sigma}\psi^j\right)\right.\nonumber\\
&+&\left. \frac{1}{4}  R_{jlmk}\eta^j\eta^l\psi^m\psi^k\right]. \label{acta}
\end{eqnarray}
 
The generalized supersymmetric BRST transformation for one dimensional
 sigma model on a curved target space is constructed by
\begin{eqnarray}
\delta_g \phi^i &=& -\psi^i \Theta[\Phi],\nonumber\\
\delta_g \psi^i &=&0,\nonumber\\
\delta_g \eta_i &=& \left(B_i -\eta_j\Gamma^j_{\ ik}\psi^k \right)\Theta[\Phi],\nonumber\\
\delta_g B_i &=& -\left(B_j\Gamma^j_{\ ik}\psi^{k} -\frac{1}{2}\eta_j R^j_{\ ilk}\psi^l\psi^k \right)\Theta[\Phi],\nonumber\\
\delta_g \Gamma^j_{\ ik} &=&\partial_m \Gamma^j_{\ ik}\psi^m \Theta[\Phi],\nonumber\\
\delta_g R^j_{\ ilk} &=&\partial_m R^j_{\ ilk}\psi^m \Theta[\Phi],
\end{eqnarray}
 where $ \Theta[\Phi]$ is the general finite field-dependent parameter. For
instance, we choose an specific  $ \Theta[\Phi]$ obtained from
the following infinitesimal field-dependent parameter using relation (\ref{fin}):
\begin{equation}
 \Theta'[\eta, \phi, B] = -  \alpha \int d\sigma\ \eta_i\left(\frac{d\phi^i}{d\sigma}-\frac{1}{2}g^{ij}B_j \right).\label{fin1}
\end{equation}
The infinitesimal change of Jacobian of the path integral measure
is calculated by exploiting relation (\ref{jac}) as 
\begin{eqnarray}
\frac{d }{d\kappa} \ln J(\kappa)&=&
 \alpha\int d\sigma \left[-B_i\frac{d\phi^i}{d\sigma} +\frac{1}{2}g^{ij}B_iB_j +
\eta_i \left( \frac{d\psi^i}{d\sigma}+\Gamma^i_{\ kj}\frac{d\phi^k}{d\sigma}\psi^j\right)\right.\nonumber\\
&-&\left. \frac{1}{4}R_{jlmk}\eta^j\eta^l\psi^m\psi^k\right].\label{ja}
\end{eqnarray}
Now, we make an ansatz for the arbitrary functional $S$ which appears in the 
expression (exponent) of the Jacobian (\ref{js}) as
\begin{eqnarray}
S  [\phi (\sigma, \kappa), \kappa ] &=&
 \int d\sigma \left[\zeta_1(\kappa)B_i\frac{d\phi^i}{d\sigma} +\zeta_2(\kappa) 
 g^{ij}B_iB_j + \zeta_3(\kappa)\eta_i \left( \frac{d\psi^i}{d\sigma}+\Gamma^i_{\ kj}\frac{d\phi^k}{d\sigma}\psi^j\right)\right.\nonumber\\
&+&\left. \zeta_4(\kappa) R_{jlmk}\eta^j\eta^l\psi^m\psi^k\right],
\end{eqnarray}
where $\zeta_1, \zeta_2, \zeta_3$ and $\zeta_4$
are $\kappa$-dependent constants which vanish at $\kappa =0$.
The  existence of the above functional is valid when it  satisfies the essential requirement
given in (\ref{mcond}) along with   (\ref{ja}). 
This leads to the following condition:
\begin{eqnarray}
&&
 \int d\sigma \left[\left(\frac{d\zeta_1 }{d\kappa} - \alpha \right)B_i\frac{d\phi^i}{d\sigma} +\left(\frac{d\zeta_2 }{d\kappa}+\frac{1}{2} \alpha \right) 
 g^{ij}B_iB_j + \left(\frac{d\zeta_3 }{d\kappa} + \alpha\right)\eta_i \left( \frac{d\psi^i}{d\sigma}+\Gamma^i_{\ kj}\frac{d\phi^k}{d\sigma}\psi^j\right)\right.\nonumber\\
&&+\left. \left(\frac{d\zeta_4 }{d\kappa} -\frac{1}{4} \alpha \right)R_{jlmk}\eta^j\eta^l\psi^m\psi^k +\left(\zeta_2 +2\zeta_4\right)
\eta_j R^j_{\ ilk}\psi^l\psi^k B^i\Theta'[\phi]\right.\nonumber\\
&&-\left.(\zeta_1 +\zeta_3)\left(B_i\frac{d\psi^i}{d\sigma}+B_j\Gamma^j_{\ ik}\psi^k \frac{d\phi^i}{d\sigma} -\frac{1}{2}\eta_jR^j_{\ ilk}\psi^l\psi^k\frac{d\phi^i}{d\sigma}
\right)\Theta' [\phi] \right] =0,
\end{eqnarray}
where we have used the antisymmetry of the Grassmann variables
and Bianchi identity of Riemann tensor.
The comparison of various terms on both sides yields the following  
constraints on the parameters   $\zeta_i (\kappa)$, where $i=1,2,3,4$ :
\begin{eqnarray}
&&\frac{d\zeta_1(\kappa)}{d\kappa} - \alpha =0,\label{x} \\
&&\frac{d\zeta_2(\kappa)}{d\kappa}+\frac{1}{2} \alpha =0, \\
&&\frac{d\zeta_3(\kappa)}{d\kappa} + \alpha =0, \\
&&\frac{d\zeta_4(\kappa)}{d\kappa} -\frac{1}{4} \alpha =0,\label{y} \\
&&  \zeta_1 (\kappa)+\zeta_3 (\kappa) =0, \label{w}\\
&& \zeta_2 (\kappa)+2\zeta_4(\kappa) =0.\label{z}
\end{eqnarray}
The solutions of the above differential equations given in (\ref{x})-(\ref{y}) are
\begin{eqnarray}
\zeta_1(\kappa) = \alpha\kappa,\ \ \zeta_2 =-\frac{1}{2} \alpha\kappa,\ \ 
\zeta_3(\kappa) =-  \alpha\kappa,\ \ \zeta_4(\kappa) = \frac{1}{4} \alpha\kappa.
\end{eqnarray}
These solutions are also consistent with relations (\ref{w}) and (\ref{z}).
Therefore, with these identifications of $\zeta_i$, action 
$S$ simplifies as
\begin{eqnarray}
S  [\phi (\sigma, \kappa), \kappa ] &=&\alpha\kappa
 \int d\sigma \left[  B_i\frac{d\phi^i}{d\sigma} -\frac{1}{2}  
 g^{ij}B_iB_j-  \eta_i \left( \frac{d\psi^i}{d\sigma}+\Gamma^i_{\ kj}\frac{d\phi^k}{d\sigma}\psi^j\right)\right.\nonumber\\
&+&\left. \frac{1}{4}  R_{jlmk}\eta^j\eta^l\psi^m\psi^k\right],
\end{eqnarray}
which vanishes at $\kappa =0$. However, at $\kappa=1$ (under generalized supersymmetry transformation),
it takes the following form
\begin{eqnarray}
S  [\phi (\sigma, 1), 1 ] &=&\alpha 
 \int d\sigma \left[  B_i\frac{d\phi^i}{d\sigma} -\frac{1}{2}  
 g^{ij}B_iB_j-  \eta_i \left( \frac{d\psi^i}{d\sigma}+\Gamma^i_{\ kj}\frac{d\phi^k}{d\sigma}\psi^j\right)\right.\nonumber\\
&+&\left. \frac{1}{4}  R_{jlmk}\eta^j\eta^l\psi^m\psi^k\right],
\end{eqnarray}
which exactly  coincides  with the effective action (\ref{acta}) for the sigma model 
on curved target space in one dimension. This shows that the
effective action for the  sigma model on curved target space emerges naturally through the Jacobian  of the
path integral measure under generalized supersymmetric transformation.
Now if we apply again the FFBRST transformation with appropriate 
choice of finite field-dependent parameter, we can get the 
sigma model in different gauges. 
 \subsection{Topological sigma model}
 In this subsection
 we discuss the topological sigma model for hyperK\"{a}hler map.
 For this purpose we start by defining  a map $\phi: {\cal M}\longrightarrow {\cal N}$ 
from a Riemannian world-manifold ${\cal M}$ to
 a Riemannian target-manifold ${\cal N}$  which deals with the homotopy classes of the
map.
This map is described by an action 
\begin{eqnarray}
S =\int_{\cal M} d^m \sigma\sqrt{g(\sigma)}g^{\alpha\beta}(\sigma)\partial_\alpha \phi^i
\partial_\beta\phi^j h_{ij}(\phi),
\end{eqnarray} 
where $m=\mbox {dim}{\cal M}$, $g^{\alpha\beta}(\sigma)$ is the metric of the world-manifold ${\cal M}$
and $h_{ij}(\phi)$ is the metric of target-manifold ${\cal N}$.
Here Greek indices $\alpha, \beta = 1,2,...,m$ denote the world indices
and indices $i,j =1,2,..., 4n$ refer to the target ones where 
$\mbox{dim} {\cal N} =4n$ is fixed. This action is   topologically  invariant under any continuous
deformation, $\phi \longrightarrow\phi +\delta\phi$, due to the large symmetry
  required by it. Therefore topological  sigma model is   intrinsically a quantum field theory. This large symmetry is BRST-quantized  \cite{bau, jmf} 
in the usual ways and the gauge is fixed by choosing suitable representatives in the homotopy classes of the maps $\phi$.

The supersymmetric BRST-quantization of the theory is achieved as follows.
First of all we introduce  topological
ghosts $\psi^i$ as well as topological antighosts $\eta^i_\alpha$ 
and
Lagrange multipliers $B^i_\alpha$ corresponding to the gauge-fixing in the theory.
Here an extra index $\alpha$ corresponds to the 
directions in the base space.
These antighosts and Lagrange multipliers
are required to satisfy the following duality condition
\begin{eqnarray}
\eta^i_\alpha -\frac{1}{3}(j_u)_\alpha^{\ \beta}\eta^j_\beta (J_u)_j^{\ i}=0,\ \ 
B^i_\alpha +\frac{1}{3}(j_u)_\alpha^{\ \beta}B^j_\beta (J_u)_j^{\ i}=0,
\end{eqnarray}
where $j_u (J_u)$  are called the almost quaternionic $(1, 1)$-tensors of ${\cal M}({\cal N})$ with $u=1,2,3$.
Now, the nilpotent supersymmetry transformations are constructed as \cite{fre}
\begin{eqnarray}
\delta \phi^i &=& -\psi^i\xi,\nonumber\\
\delta \psi^i &=& 0,\nonumber\\
\delta \eta^i_\alpha &=& B^i_\alpha \xi-\Gamma^i_{\ jk}\psi^j\eta_\alpha^k \xi
-\frac{1}{4}(j_u)_\alpha^{\ \beta}D_k (J_u)_j^{\ i}\psi^k\eta^{\ j}_\beta\xi,\nonumber\\ 
\delta B^i_\alpha &=& -\frac{1}{2} R_{jk\ l}^{\ \ i}\psi^j\psi^k\eta_\alpha^l \xi  + \Gamma^i_{\ jk}\psi^j B^k_\alpha  \xi+\frac{1}{4}(j_u)_\alpha^{\ \beta}D_k(J_u)_j^{\ i}\psi^k B^j_{\beta}\xi \nonumber\\
&-& \frac{1}{4}(j_u)_\alpha^{\ \beta}D_mD_k(J_u)_j^{\ i}\psi^m\psi^k \eta^j_{\beta} \xi + \frac{1}{16}D_k (J_u)_j^{\ i}D_l (J_u)_m ^{\ j}\psi^k\psi^l \eta^m_\alpha\xi \nonumber\\
&-&\frac{1}{16}\epsilon_{uvz}(j_z)_\alpha^{\ \beta}D_k(J_u)_j^{\ i}D_l (J_v)_m ^{\ j}\psi^k\psi^l\eta^m_\beta\xi, \label{super}
\end{eqnarray}
where $\xi$ is global anticommuting parameter. Here the covariant derivative of $\psi^i$
is defined by
\begin{equation}
D_\alpha\psi^i =\partial_\alpha\psi^i +\Gamma^i_{\ jk}\partial_\alpha\phi^j\psi^k.
\end{equation}
Now, with these introductions the supersymmetric action for
topological sigma model is constructed by \cite{fre}
\begin{eqnarray}
S=S_{bose}+S_{fermi},\label{acta2}
\end{eqnarray}
where  
\begin{eqnarray}
S_{bose}& =&\int_{\cal M} d^m\sigma \sqrt{g} g^{\alpha\beta}h_{ij}B^i_\alpha
\left(\partial_\beta\phi^j -\frac{1}{8}B^j_\beta\right),
\nonumber\\
S_{fermi} &=&\int_{\cal M} d^m\sigma \sqrt{g}\left[  -g^{\alpha\beta}h_{ij}
\eta^i_\alpha D_\beta\psi^j +\frac{1}{16}R_{ijkl}g^{\alpha\beta}\eta_\alpha^i\eta^j_\beta\psi^k\psi^l\right.\nonumber\\
&+&\left.
 \frac{1}{4}\eta_\alpha^m (j_u)^{\beta\alpha} D_k (J_u)_{mj}\partial_\beta\phi^j\psi^k 
+\frac{1}{32}\eta^i_\alpha \eta^l_\beta (j_u)^{\alpha\beta}D_mD_k (J_u)_{li}\psi^m\psi^k 
\right.\nonumber\\
&-&\left.   \frac{1}{128}g^{\alpha\beta}\eta^i_\alpha\eta_\beta^m D_k
(J_u)_{li}D_n(J_u)_m^{\ l}\psi^k\psi^n\right.\nonumber\\
&+&\left. \frac{1}{128}\eta_\alpha^i\eta_\beta^m \epsilon_{uvz}(j_z)^{\alpha\beta}
D_k (J_u)_{li}D_n (J_v)_m^{\ l}\psi^k\psi^n \right].
\end{eqnarray}
which remains invariant under the supersymmetry transformations given in (\ref{super}).

The supersymmetry of topological sigma model given in  (\ref{super}) is generalized as 
\begin{eqnarray}
\delta \phi^i &=& -\psi^i\Theta[\phi],\nonumber\\
\delta \psi^i &=& 0,\nonumber\\
\delta \eta^i_\alpha &=& B^i_\alpha \Theta[\phi]-\Gamma^i_{\ jk}\psi^j\eta_\alpha^k \Theta[\phi]
-\frac{1}{4}(j_u)_\alpha^{\ \beta}D_k (J_u)_j^{\ i}\psi^k\eta^{\ j}_\beta\Theta[\phi],\nonumber\\ 
\delta B^i_\alpha &=& -\frac{1}{2} R_{jk\ l}^{\ \ i}\psi^j\psi^k\eta_\alpha^l \Theta[\phi] + \Gamma^i_{\ jk}\psi^j B^k_\alpha\Theta[\phi] +\frac{1}{4}(j_u)_\alpha^{\ \beta}D_k(J_u)_j^{\ i}\psi^k B^j_{\beta}\Theta[\phi]\nonumber\\
&-& \frac{1}{4}(j_u)_\alpha^{\ \beta}D_mD_k(J_u)_j^{\ i}\psi^m\psi^k \eta^j_{\beta}\Theta[\phi]+ \frac{1}{16}D_k (J_u)_j^{\ i}D_l (J_u)_m ^{\ j}\psi^k\psi^l \eta^m_\alpha\Theta[\phi]\nonumber\\
&-&\frac{1}{16}\epsilon_{uvz}(j_z)_\alpha^{\ \beta}D_k(J_u)_j^{\ i}D_l (J_v)_m ^{\ j}\psi^k\psi^l\eta^m_\beta\Theta[\phi],
\end{eqnarray}
 where $ \Theta[\Phi]$ is an arbitrary finite field-dependent parameter.
However it can be specified to have some particular values. For example, we
choose the $\Theta[\phi]$ obtained from
the following infinitesimal field-dependent parameter using relation (\ref{fin}):
\begin{equation}
 \Theta'[\eta, \phi, B] = -   \int_{\cal M} d^m\sigma\ 
 \sqrt{g}g^{\alpha\beta}h_{ij}\eta^i_\alpha\left(\partial_\beta\phi^j -\frac{1}{8} 
 B^j_\beta \right).\label{fin2}
\end{equation}
Now, exploiting relation (\ref{jac}), the infinitesimal change of Jacobian of 
the path integral measure
is calculated as 
\begin{eqnarray}
\frac{d }{d\kappa} \ln J(\kappa)&=&
\int_{\cal M} d^m\sigma \sqrt{g}\left[ -g^{\alpha\beta}h_{ij}B^i_\alpha
\left(\partial_\beta\phi^j -\frac{1}{8}B^j_\beta\right) +g^{\alpha\beta}h_{ij}
\eta^i_\alpha D_\beta\psi^j \right.\nonumber\\
&-&\left.\frac{1}{16}R_{ijkl}g^{\alpha\beta}\eta_\alpha^i\eta^j_\beta\psi^k\psi^l
-\frac{1}{4}\eta_\alpha^m (j_u)^{\beta\alpha} D_k (J_u)_{mj}\partial_\beta\phi^j\psi^k 
\right.\nonumber\\
&-&\left. \frac{1}{32}\eta^i_\alpha \eta^l_\beta (j_u)^{\alpha\beta}D_mD_k (J_u)_{li}\psi^m\psi^k +\frac{1}{128}g^{\alpha\beta}\eta^i_\alpha\eta_\beta^m D_k
(J_u)_{li}D_n(J_u)_m^{\ l}\psi^k\psi^n\right.\nonumber\\
&-&\left. \frac{1}{128}\eta_\alpha^i\eta_\beta^m \epsilon_{uvz}(j_z)^{\alpha\beta}
D_k (J_u)_{li}D_n (J_v)_m^{\ l}\psi^k\psi^n\right]. \label{inf}
\end{eqnarray}
Further, we make an arbitrary ansatz for the functional $S[\Phi]$ (\ref{js})
having similar terms as in RHS of (\ref{inf}). Henceforth, $S[\Phi]$ is defined by  
\begin{eqnarray}
S[\Phi(\sigma, \kappa), \kappa ]& =&\int_{\cal M} d^m\sigma  \left[\zeta_1 (\kappa)g^{\alpha\beta}h_{ij}B^i_\alpha
 \partial_\beta\phi^j +\zeta_2 (\kappa)g^{\alpha\beta}h_{ij}B^i_\alpha B^j_\beta  +\zeta_3 (\kappa)g^{\alpha\beta}h_{ij}
\eta^i_\alpha D_\beta\psi^j \right.\nonumber\\
&+&\left.\zeta_4 (\kappa)R_{ijkl}g^{\alpha\beta}\eta_\alpha^i\eta^j_\beta\psi^k\psi^l
+\zeta_5 (\kappa)\eta_\alpha^m (j_u)^{\beta\alpha} D_k (J_u)_{mj}\partial_\beta\phi^j\psi^k 
\right.\nonumber\\
&+&\left. \zeta_6 (\kappa)\eta^i_\alpha \eta^l_\beta (j_u)^{\alpha\beta}D_mD_k (J_u)_{li}\psi^m\psi^k +\zeta_7 (\kappa)g^{\alpha\beta}\eta^i_\alpha\eta_\beta^m D_k
(J_u)_{li}D_n(J_u)_m^{\ l}\psi^k\psi^n\right.\nonumber\\
&+&\left. \zeta_8 (\kappa)\eta_\alpha^i\eta_\beta^m \epsilon_{uvz}(j_z)^{\alpha\beta}
D_k (J_u)_{li}D_n (J_v)_m^{\ l}\psi^k\psi^n\right], \label{s1}
\end{eqnarray}
where $\zeta_i (\kappa), i=1,2,...,8,$ are $\kappa$-dependent constants
satisfying initial boundary conditions.
The equations (\ref{inf}) and (\ref{s1}) together with condition
(\ref{mcond}) yield the following differential equations
\begin{eqnarray}
&&\frac{d\zeta_1(\kappa)}{d\kappa} - \sqrt{g} =0, \\
&&\frac{d\zeta_2(\kappa)}{d\kappa}+\frac{1}{8} \sqrt{g}=0, \\
&&\frac{d\zeta_3(\kappa)}{d\kappa} + \sqrt{g}=0, \\
&&\frac{d\zeta_4(\kappa)}{d\kappa} -\frac{1}{16} \sqrt{g}  =0, \\
&&\frac{d\zeta_5(\kappa)}{d\kappa} - \frac{1}{4}\sqrt{g} =0, \\
&&\frac{d\zeta_6(\kappa)}{d\kappa}-\frac{1}{32}\sqrt{g}=0, \\
&&\frac{d\zeta_7(\kappa)}{d\kappa} + \frac{1}{128}\sqrt{g} =0, \\
&&\frac{d\zeta_8(\kappa)}{d\kappa} -\frac{1}{128} \sqrt{g} =0. \\
\end{eqnarray}
The above linear differential equations are exactly solvable.
Their solutions  satisfying the
initial conditions $\xi_i(\kappa =0) =0, i=1,2,3,4$ are
\begin{eqnarray}
\zeta_1(\kappa) &=&\sqrt{g}\kappa,\ \ \zeta_2(\kappa) =-\frac{1}{8}\sqrt{g}\kappa,\ \ \zeta_3(\kappa) =-\sqrt{g}\kappa,\ \ \zeta_4(\kappa) =\frac{1}{16}\sqrt{g}\kappa,\nonumber\\
\zeta_5(\kappa) &=&\frac{1}{4}\sqrt{g}\kappa,\ \ \zeta_6(\kappa) =\frac{1}{32}\sqrt{g}\kappa,\ \ \zeta_7(\kappa) =- \frac{1}{128}\sqrt{g}\kappa,\ \ \zeta_8(\kappa) =\frac{1}{128} \sqrt{g}\kappa.
\end{eqnarray}
 With these values of constants, the functional $S[\phi(\sigma, \kappa)]$ reduces 
 to
 \begin{eqnarray}
S[\Phi(\sigma, \kappa), \kappa]& =&\kappa \int_{\cal M} d^m\sigma \sqrt{g}\left[g^{\alpha\beta}h_{ij}B^i_\alpha
\left(\partial_\beta\phi^j -\frac{1}{8}B^j_\beta\right) -g^{\alpha\beta}h_{ij}
\eta^i_\alpha D_\beta\psi^j \right.\nonumber\\
&+&\left.\frac{1}{16}R_{ijkl}g^{\alpha\beta}\eta_\alpha^i\eta^j_\beta\psi^k\psi^l
+\frac{1}{4}\eta_\alpha^m (j_u)^{\beta\alpha} D_k (J_u)_{mj}\partial_\beta\phi^j\psi^k 
\right.\nonumber\\
&+&\left. \frac{1}{32}\eta^i_\alpha \eta^l_\beta (j_u)^{\alpha\beta}D_mD_k (J_u)_{li}\psi^m\psi^k -\frac{1}{128}g^{\alpha\beta}\eta^i_\alpha\eta_\beta^m D_k
(J_u)_{li}D_n(J_u)_m^{\ l}\psi^k\psi^n\right.\nonumber\\
&+&\left. \frac{1}{128}\eta_\alpha^i\eta_\beta^m \epsilon_{uvz}(j_z)^{\alpha\beta}
D_k (J_u)_{li}D_n (J_v)_m^{\ l}\psi^k\psi^n\right],
\end{eqnarray}
which vanishes at $\kappa=0$. However, for $\kappa =1$,
it becomes 
\begin{eqnarray}
S[\Phi(\sigma,1), 1]& =&\int_{\cal M} d^m\sigma \sqrt{g}\left[g^{\alpha\beta}h_{ij}B^i_\alpha
\left(\partial_\beta\phi^j -\frac{1}{8}B^j_\beta\right) -g^{\alpha\beta}h_{ij}
\eta^i_\alpha D_\beta\psi^j \right.\nonumber\\
&+&\left.\frac{1}{16}R_{ijkl}g^{\alpha\beta}\eta_\alpha^i\eta^j_\beta\psi^k\psi^l
+\frac{1}{4}\eta_\alpha^m (j_u)^{\beta\alpha} D_k (J_u)_{mj}\partial_\beta\phi^j\psi^k 
\right.\nonumber\\
&+&\left. \frac{1}{32}\eta^i_\alpha \eta^l_\beta (j_u)^{\alpha\beta}D_mD_k (J_u)_{li}\psi^m\psi^k -\frac{1}{128}g^{\alpha\beta}\eta^i_\alpha\eta_\beta^m D_k
(J_u)_{li}D_n(J_u)_m^{\ l}\psi^k\psi^n\right.\nonumber\\
&+&\left. \frac{1}{128}\eta_\alpha^i\eta_\beta^m \epsilon_{uvz}(j_z)^{\alpha\beta}
D_k (J_u)_{li}D_n (J_v)_m^{\ l}\psi^k\psi^n\right],
\end{eqnarray}
which is the exact expression of the supersymmetric topological
sigma model (\ref{acta2}) in $m$-dimensions. Therefore, we generated the effective action for
supersymmetric topological sigma model by calculating
 the Jacobian of the path integral under generalized supersymmetry transformations with 
 appropriate transformation parameter.
 Further, we observe that under further generalized supersymmetry 
 with appropriate field-dependent parameter we can map the topological sigma model
 from one gauge to another.
  \section{Conclusions}
In this paper, we have described the
mechanism of generalized BRST transformation
to establish the connection between two different gauges of
Maxwell theory. In the same fashion, we have proposed the idea behind generalizing
supersymmetry. We have generalized the BRST 
supersymmetry by allowing the
transformation parameter to be finite and field-dependent.
The generalized supersymmetry retains the invariance at
the level of the action only, however, the generating functional 
does not. 
The obvious reason for this is that the path integral measure
is not invariant under the transformation.
We have shown that under such generalized supersymmetry the path integral measure of functional
integral changes non-trivially. We have sketched a novel feature originating from such
 non-trivial Jacobian under   generalized
supersymmetry.  With suitable 
choices of finite and field-dependent transformation parameters, the Jacobian
generates the supersymmetric actions corresponding to
sigma models. In fact the Jacobian reproduces the well known supersymmetric actions of sigma models.

It is useful to note that not all supersymmetric actions may be generated in this manner. As discussed earlier, only those
actions that are BRST exact may be obtained. This is essentially tied to the fact that the jacobian of FFBRST transformation
is BRST exact.

The present analysis highlights the important role of symmetry in the abstraction of supersymmetric actions. As is 
well known, for nonsupersymmetric theories, gauge invariance is crucial for the obtention of actions. The calculation
of the one loop effective action using a gauge invariant regularisation provides a striking example in this context.
In (1+1) dimensions the computation can be done exactly and yields the Schwinger model. In (2+1) (or higher) dimensions
the result cannot be obtained exactly and one takes recourse to the derivative expansion. The first term 
is the (single derivative) Chern-Simons term, the second is the (two derivative) Maxwell term and so on. For supersymmetric
theories, gauge invariance gets replaced by BRST invariance. Our analysis shows in a precise way the role of this invariance 
in obtaining a certain class of supersymmetric actions. This was explicitly shown for two supersymmetric sigma models- a sigma model in one dimension and a topological sigma model in general dimensions. It illustrates the robustness of the technique in the sense 
that it may be applied in quite distinct situations.
We note that
under the action of further 
generalized supersymmetry transformations with appropriate 
transformation parameters we will be able to
connect the supersymmetric sigma models in different gauges, exactly as was discussed 
for the Maxwell theory. We hope this formulation will help to systematically
construct the supersymmetric actions for
sigma models in an elegant manner as well as provides a deeper understanding.

 Let us discuss the arbitrariness in this scheme. It is contained in
the choice of the FFBRST parameter $\Theta$ defined generally in (\ref{fin})
and in (\ref{fin1}) and (\ref{fin2}) for the specific models under consideration. Once
this choice is made the rest follows systematically. The point is that
once $\Theta$ is defined, the infinitesimal change of the Jacobian is
calculated from the specified formula (\ref{js}). The structure of this
change determines the ansatz to be adopted for the functional that
appears in the exponent of the Jacobian (\ref{jacob}). This eventually yields
the final answer. The choice of the $\Theta$ parameter is somewhat
akin to choosing a good gauge. A judicious choice of this parameter is
important to get meaningful results.

\end{document}